\documentclass[aps,amsfonts,amsmath,amssymb,graphics,floatfix,epsfig,twocolumn,fleqn,balancelastpage]{revtex4}
\usepackage[dvips]{graphicx}
\usepackage{graphics}
\usepackage{epsfig}
\begin{document}
\title{On the theory underlying the Car-Parrinello method and the role of the fictitious mass parameter.\footnote{The following article has been accepted by The Journal of Chemical Physics. After it is published, it will be found at \url{http://jcp.aip.org/}}}
\author{Paul Tangney}
\altaffiliation[Current address: ]
{The Molecular Foundry, Lawrence Berkeley National Laboratory, CA 94720}
\email{pttangney@lbl.gov}
\date{\today}
\affiliation{ International School for Advanced Studies, via Beirut 2-4, 34013 Trieste, Italy\\
The Molecular Foundry, Lawrence Berkeley National Laboratory, CA 94720\\
Department of Physics, University of California, Berkeley, CA 94720}
\begin{abstract}
The theory underlying the Car-Parrinello extended-lagrangian approach to
{\em ab initio} molecular dynamics (CPMD) is reviewed and reexamined using 
 'heavy' ice as a test system. 
It is emphasized that the adiabatic decoupling in CPMD is not a decoupling of 
electronic orbitals from the ions but only a decoupling of a subset
of the orbital vibrational modes from the rest of the necessarily-coupled 
system of orbitals and ions. 
Recent work ( J. Chem. Phys. {\bf 116} , 14 (2002) ) has pointed out that, 
due to the orbital-ion coupling that remains
once adiabatic-decoupling has been achieved, a large
value of the fictitious mass $\mu$ can lead to systematic errors in the 
computed forces in CPMD.
These errors are further investigated in the present work with a focus on those parts 
of these errors that are not corrected simply by rescaling the masses of the ions.
It is suggested that any comparison of the efficiencies of Born-Oppenheimer 
molecular dynamics (BOMD) and CPMD should be performed at a similar level of accuracy.
If accuracy is judged according to the average magnitude of the systematic errors
in the computed forces, the efficiency of BOMD compares more favorably to that of CPMD
than previous comparisons have suggested.
\end{abstract}
\maketitle
\section{Introduction} \label{section:intro}
Since its introduction by Car and Parrinello in 1985\cite{carpar}, the field of 
{\em ab initio} molecular dynamics (AIMD) has grown very rapidly.
Although many of its initial successes 
involved its application to questions in condensed matter physics and materials science, 
it has now been applied with success to a wide range of problems in 
diverse areas of science such as chemistry, biology and geophysics
and contributed a great deal to our understanding of these fields.
The central idea of the method is that the Kohn-Sham orbitals\cite{kohnsham} that 
describe the electronic state of a system within density functional
theory\cite{hohenbergkohn} (DFT) are evolved simultaneously with the ions
as classical degrees of freedom.
In the original approach of Car and Parrinello, an inertial parameter is introduced in 
order to associate a Newtonian dynamics with the electronic orbitals. 
This parameter is an unphysical quantity that is commonly referred to as the fictitious mass, $\mu$.  
While it is known that the introduction of this extra inertia into
a system affects its dynamics, the extent to which dynamics are altered
and the way in which they are altered are only beginning to be studied in 
detail\cite{us,grossman,schwegler}.

In this paper some of the theoretical ideas that underlie the original,
and still widely-used, Car-Parrinello approach to {\em ab initio} molecular dynamics 
are examined and the possible importance of the 
fictitious mass effects are investigated by studying the example of ice. 
In what follows, this approach to AIMD 
will be referred to as Car-Parrinello molecular 
dynamics (CPMD). In CPMD the equations of motion of the ions (with coordinates $R_{I}$ and
masses $M_I$) and
the electronic orbitals $|\psi_i \rangle$ are derived from the classical lagrangian 
\begin{equation}
L = \mu \sum_{i}\langle\dot{\psi}_{i}|\dot{\psi}_{i}\rangle
+ \frac{1}{2}\sum_{I}M_{I}\dot{\mathbf{R}}_{I}^{2} 
- E[\{\psi_{i}\},\{\mathbf{R}_{I}\}]
\label{eqn:lagrangian}
\end{equation}
and the motion is further constrained to the hypersurface defined by the orbital
orthogonality condition $\langle \psi_i | \psi_j \rangle = \delta_{ij} ; \forall i,j$\cite{ortho}.
Inspired by the ideas of Car and Parrinello, an alternative approach to AIMD
was proposed\cite{payne} in which the electronic orbitals are kept as close as
possible to the instantaneous ground state using a combination of extrapolation 
from earlier times and explicit minimization. This general approach will be
referred to here as Born-Oppenheimer molecular dynamics (BOMD).

It has been known since its introduction that CPMD is an approximation to a fully
converged BOMD simulation in which the electronic system is exactly in its 
ground state at every instant of the ionic motion, however CPMD is frequently
preferred over BOMD because it is deemed computationally more efficient.
In CPMD simulations, because the orbitals are treated as classical degrees 
of freedom on the same footing as the ions, there exists an energy (which includes the 
kinetic energy associated with the fictitious classical motion of the orbitals) 
that is perfectly conserved as long as the time step that is used is small enough 
to adequately integrate the equations of motion.
On the other hand, while the conserved quantity in BOMD is the physically
meaningful total energy of the system of electrons and ions, its conservation
in practice is always imperfect. There always exists a drift in the calculated
total energy of a BOMD simulation because the electronic orbitals are never perfectly
converged to the ground state. While the magnitude of this drift can systematically 
be reduced by more fully converging the minimization of the electronic system at each 
time step, any such improvement in accuracy must always be balanced by the computational
expense involved.
The argument has frequently been made\cite{marx,kuo,caryip}, that for many systems CPMD is computationally 
superior to BOMD because the computational overhead required to make BOMD conservative within acceptable 
limits is greater than that required to perform a perfectly conservative CPMD simulations with
a value of $\mu$ chosen in line with conventional wisdom\cite{pastore,marx,grossman}.
Since BOMD can be as fast as, or faster than, CPMD depending on the size of the energy 
drift that is tolerated, an implicit assumption of this argument is that BOMD can only
match the accuracy of CPMD when this drift of energy is very small.
In this paper, this assumption is questioned by pointing out that, for commonly
used values of the fictitious mass $\mu$, there can exist large errors in the
forces on the ions in a perfectly conservative and well-controlled CPMD simulation.

The emphasis of this paper is on the microscopic details of CPMD simulations, i.e. on the
ability of the CPMD approach to evolving the electronic orbitals in time to produce (on average) the 
correct ground state forces on the ions. The important question of how deviations of forces from the correct forces 
relate to possible deviations in thermodynamic quantities from those that would be obtained using the correct forces is not 
addressed in the present work. 

Car and Parrinello introduced their method in the context of coupling Kohn-Sham density functional 
theory\cite{kohnsham,hohenbergkohn} to molecular dynamics, however, it has since been applied successfully 
to other electronic structure methods\cite{carter}. It has also been applied in a purely
classical context to evolve induced dipoles\cite{sprik_klein,sprik} and fluctuating charges\cite{rick_berne} 
in molecular dynamics simulations. None of these applications will be discussed in the present
work but, clearly,  many of the concepts discussed here have analogies within these methods.

\section{Background} \label{section:background}
It has been argued, and demonstrated for the case of silicon\cite{pastore,marx}, that 
although there exist small instantaneous deviations of the forces in a CPMD simulation from
the ground state forces at the same ionic positions, these deviations average
to close to zero on a femtosecond time scale and therefore do not result in serious  
errors in thermodynamic properties\cite{pastore,carlouie,marx,caryip}.
The explanation proposed for this behaviour\cite{pastore} is that the electronic 
orbitals perform small-amplitude ultra high frequency oscillations around an equilibrium
that slowly evolves as the ions move. If this equilibrium corresponds to the electronic 
ground state, and the amplitudes of the oscillations are small, the forces on the ions 
oscillate about the true ground state forces and errors in the forces average to zero 
on time scales that are relevant to ion dynamics.

The fact that a dynamics is associated with electronic orbitals means
that, according to the equipartition theorem, the subsystems of orbitals and ions 
should equilibrate with one another and reach a common temperature. 
It has been observed, however, that for systems in which there is a substantial gap 
in the Kohn-Sham energy spectrum between occupied and unoccupied electronic states, 
there is no systematic net transfer of kinetic energy between orbitals and ions and
therefore the orbitals remain at a temperature that is very low relative to the
ionic temperature. This low temperature means that orbitals do 
not have the kinetic energy required to leave the energy basin of the 
electronic ground state. 

The non-equilibration of orbitals and ions has been explained\cite{pastore} as being 
due to a gap in the vibrational frequency spectrum of the coupled orbital-ion system 
that separates the ultra high frequency orbital modes from all the lower frequency 
modes of the system.  The ultra high frequency oscillations of the orbitals 
are ``adiabatically decoupled'' from the rest of the system.

While the arguments presented above appear to work well for
silicon\cite{pastore,us}, little work has been done to 
verify that CPMD forces average to the ground state forces for other systems. 
Furthermore, some important details of the arguments of Pastore {\em et al.} have 
received little attention: Adiabatic decoupling is {\em not} a decoupling of orbitals from ions. 
Rather, it is only a decoupling of the quasi-independent orbital modes that consist of 
ultra high frequency oscillations of the orbitals from the lower frequency modes of the system.  
The lower frequency modes of the system generally involve both ions and orbitals. 
The slow (ionic time scale) component of the orbital motion is frequently
completely neglected in discussions of the CPMD method\cite{grossman, caryip}, 
however, and it is often stated that the lowest frequency motion in the orbital 
subsystem is approximately related to the Kohn-Sham energy gap of the system $E_g$ 
through \cite{vandevondele2,marx,grossman,caryip}
\begin{equation}
\;\;\;\;\;\;\;\;\;\;\;\;\;\;\;\;\;\;\;\;\;\;\;\;\;\;\;\;\;\; \omega \approx \sqrt{2 E_g/\mu}
\label{eqn:minfreq}
\end{equation}.
In an insulator, this frequency is very high compared to typical ionic frequencies.
Clearly, this view is not compatible with the electronic orbitals remaining at or 
near the electronic ground state because the ground state {\em by definition} evolves 
on ionic time scales. 

The slow component of orbital motion due to the evolution of the ground state 
is always present and, as a result, there is always a continuous exchange of 
energy and momentum between orbitals and ions. Furthermore, for many systems, 
this ionic time scale component of the orbital motion has been found to be 
appreciable\cite{blochl_and_parrinello,us,blochl2}.
It will be demonstrated in section \ref{section:timescale} of the present work 
that this ionic-time scale contribution to the motion of the orbitals actually 
accounts for the vast majority of their 
classical ``fictitious'' kinetic energy in the simulations of ice reported here.

The argument of Pastore {\em et al.}\cite{pastore} that forces in CPMD oscillate about 
the true ground state forces relies on the assumption that the ultra high frequency oscillations 
of the orbitals are about the electronic ground state, however, for any finite $\mu$ it will be shown
in section \ref{section:orbitals} that this is not the case. As a consequence of the 
ionic time scale motion of the electronic orbitals in CPMD, 
the orbitals do not oscillate around the ground state but about average values that are displaced 
from the ground state. 
Furthermore, although the resulting errors in the forces are small for materials in which
valence electrons are delocalized and therefore have a low quantum kinetic energy
(such as silicon), for other materials the forces can deviate strongly from their ground state
values if commonly used values of the fictitious mass parameter are used\cite{us}, however, 
these errors can systematically be reduced by reducing the magnitude of $\mu$.

It has previously been proposed that an effect of the orbital-ion coupling is simply
to rescale the masses of the ions\cite{blochl_paw,us,blochl2}, an effect that does not alter 
thermodynamic properties, however, theoretically, the error induced by orbital-ion coupling only
reduces to a mass correction in the limit in which ions do not interact with one another, i.e.
in the limit of an infinitely dilute gas\cite{blochl2}. For condensed systems most, but not all,
of the error can be corrected by simply rescaling the masses of the ions.
One goal of the present paper is to review, to extend, and to help clarify some of the analysis
of reference \onlinecite{us}. Another goal is to stimulate further investigation by using
`heavy'  ice as a test system to highlight the importance of the effects
described in reference \onlinecite{us} with particular emphasis on the part of the errors
in the forces that is neither oscillatory on a very short time scale, or equivalent to 
a simple renormalization of the masses of the ions.
It is shown that, for commonly used values of $\mu$, the magnitude of this part 
of the force errors is much larger than errors in the forces in typical BOMD simulations.

Water is arguably the most important system studied with {\em ab initio} molecular dynamics.
CPMD has been applied extensively to the study of liquid water\cite{grossman,schwegler,kuo,water} and to the study 
of biological systems where water is present. 
The water molecules in ice have much the same properties as they do in liquid water, 
but ice has a larger electronic energy gap making it easier to simulate with CPMD. 
The goal here is to investigate the accuracy of the bare Car-Parrinello method without 
any further sources of error. In simulations of liquid water the
smaller energy gap between occupied and unoccupied states and the 
high frequency of the O-H stretch vibrational mode can result in
a continuous drift of kinetic energy into the orbital subsystem that 
would further complicate our analysis\cite{grossman}. 
These problems are avoided here by studying `heavy' ice rather than water.
The larger energy gap of ice and the lower frequency of the O-D stretch
in D$_2$O ensures that no such coupling occurs in the simulations reported here.

The fact that the water molecules are confined to a lattice means that 
large deviations of the electronic structure of the individual molecules from their 
average electronic structure are less likely to occur in ice than in liquid water.
As mentioned above, a rigid electronic structure 
means that errors associated with $\mu$ only affect 
dynamical properties whereas changes in the local electronic 
structure due to interactions between molecules may lead to 
errors in thermodynamic properties.
For these reasons, the errors in CPMD reported here for ice are 
taken to be indicative of similar or more serious problems in liquid water. 

The fictitious mass effects described here and in reference \onlinecite{us} should not be 
confused with those described by Grossman {\em et al.}\cite{grossman}.
The work of Grossman {\em et al.} pointed out some problems that occurred in simulations of 
 water due to the adiabatic decoupling condition breaking down and energy passing continuously into 
the ultra high frequency orbital modes from the lower frequency modes. 
The present work is only concerned with fictitious mass dependent errors that are present in a 
well-performed Car-Parrinello simulation where the adiabatic decoupling condition is perfectly 
maintained and therefore the ultra high frequency orbital modes are energetically isolated.

\section{Theory} \label{section:theory}
In this section the dynamics of the orbitals and the ions
that result from the Car-Parrinello lagrangian of equation 
\ref{eqn:lagrangian} are analysed in order to gain a better
understanding of the different contributions to the errors in the forces on the ions.

The equations of motion of the orbitals and ions that are derived from equation 
\ref{eqn:lagrangian} are
\begin{equation}
\;\;\;\;\;\;\;\;\;\;\;\;\;\;\;\;\;\;\;\mu\ddot{\psi}_{i}(\mathbf{r}) = 
-\frac{\delta E[\{\psi_{i}\},\{\mathbf{R}_{I}\}]}{\delta 
\psi_{i}^{*}(\mathbf{r})} 
\label{eqn:electrons1}
\end{equation}
\begin{equation}
\;\;\;\;\;\;\;\;\;\;\;\;\;\;\;\;\;\;\;\textrm{M}_{I}\ddot{R}_{I}^{\alpha} = 
-\frac{ \partial E [\{\psi_{i}\},\{\mathbf{R}_{I}\}]}
{\partial R_{I}^{\alpha}} 
\label{eqn:ions1}
\end{equation}

where the orbital orthonormality condition is implicit in the 
functional derivatives $\delta / \delta \psi_{i}^{*}(\mathbf{r})$ and all functional derivatives
throughout this paper. The notation
$\partial/\partial R_{I}^{\alpha}$ indicates that a partial derivative is performed
with respect to $R_{I}^{\alpha}$ with all other ion Cartesian coordinates fixed and with
the orbitals $\psi_i$ fixed. In the remainder of the paper
it will also be necessary to use the notation
$\nabla_{I}^{\alpha}$ to denote a partial derivative with respect to 
$R_{I}^{\alpha}$ with all other ion Cartesian coordinates fixed 
{\em but not with the orbitals fixed}. 

Equation~\ref{eqn:ions1} produces the correct Born-Oppenheimer dynamics if the set of orbitals 
$\{ \psi_i\}$ at which the derivatives $\frac{ \partial E [\{\psi_{i}\},\{\mathbf{R}_{I}\}]}
{\partial R_{I}^{\alpha}}$ are evaluated is the set of ground state orbitals $\{ \psi_{i}^{g.s.}\}$, i.e.
 the set of orbitals that minimize the Kohn-Sham energy functional $E [\{\psi_{i}\},\{\mathbf{R}_{I}\}]$
for fixed ionic positions $\{\mathbf{R}_{I}\}$.
The quality of the forces on the ions therefore depends on the ability of 
equation \ref{eqn:electrons1} to produce a dynamics for the orbitals that keeps them close to this 
instantaneous ground state as the ions move.
In section \ref{section:orbitals}, therefore, the dynamics of the orbitals are analysed in the regime of 
small deviations from the ground state in order to gain a better understanding of the different
contributions to the errors in the forces on the ions. In section \ref{section:ions} the effects of the 
orbital dynamics on the forces on the ions are discussed.

\subsection{The dynamics of the orbitals} \label{section:orbitals}

As stated previously, a common view of Car-Parrinello dynamics is that the Car-Parrinello orbitals $\{\psi_i\}$
oscillate rapidly around the electronic ground state so that deviations from the
ground state average to close to zero on the longer time scales that are relevant to ionic dynamics.
In this section it is demonstrated that this is not the case if $\mu$ is large, 
and an expression for the average value of the electronic orbitals is derived.

For a given set of ionic coordinates $\{\mathbf{R}_{I}\}$, the electronic ground state is well defined.
It is therefore convenient to introduce the quantity $|\delta \psi_i\rangle \equiv |\psi_i\rangle 
- |\psi_{i}^{g.s.}\rangle$, i.e. the deviation of the $i$th Car-Parrinello orbital $|\psi_i\rangle$ 
from its ground state value.
The state of the system at a given instant of a Car-Parrinello dynamics can be completely specified by specifying the 
values of the dynamical variables $\{\mathbf{R}_{I}\}, \{ \delta \psi_{i}\} $ and their time derivatives 
$\{\mathbf{\dot{R}}_{I}\}, \{ \delta \dot{\psi}_{i}\}$ at that instant. 
In what follows, the dynamics of the $\{ \delta \psi_{i}\} $ will be analyzed. 
An assumption that will be made
throughout the analysis is that the $\{ \delta \psi_{i}\} $ are small enough that certain
$\psi$-dependent quantities are well described by truncations at linear order in $\delta \psi$ 
of Taylor expansions about the electronic ground state.

A Taylor expansion of $\delta E / \delta \psi_{i}^{*}(\mathbf{r})$ about the ground state gives
\begin{eqnarray}
\frac{\delta E}{\delta \psi_{i}^{*}(\mathbf{r})}  & = &
\frac{\delta E}{\delta \psi_{i}^{*}({\mathbf r})} \Bigg |_{g.s.}  
  +  \sum_j \int \Bigg [ \frac{\delta^2 E}{\delta \psi_{j}^{*}(\mathbf{r'}) 
\delta \psi_{i}^{*}(\mathbf{r})}\Bigg|_{g.s.} \delta \psi_{j}^{*}(\mathbf{r'}) \nonumber \\
 & + & \frac{\delta^2 E}{\delta \psi_{j}(\mathbf{r'}) 
\delta \psi_{i}^{*}(\mathbf{r})}\Bigg|_{g.s.} \delta \psi_{j}(\mathbf{r'})\Bigg ] d\mathbf{r'} +
\mathcal{O}(\delta \psi^2) \label{eqn:taylor}
\end{eqnarray}
where the first term on the right hand side vanishes by definition of the ground state.

The second time derivative of $\psi_{i}(\mathbf{r})$ may be written as
\begin{eqnarray}
\;\;\;\;\;\;\;\;\;\;\;\;\;\;\;\;\;\;\;\ddot{\psi}_i(\mathbf{r}) = \delta \ddot{\psi}_i(\mathbf{r}) + \ddot \psi_i^{g.s.}(\mathbf{r}) \label{eqn:orbacc}
\end{eqnarray}
where
\begin{equation}
\ddot \psi_i^{g.s.}(\mathbf{r})  =   \sum_{I,\alpha}\ddot{R}_{I}^{\alpha}
\nabla_{I}^{\alpha}\psi_{i}^{g.s.}(\mathbf{r}) 
+\sum_{I,\alpha,J,\beta} \dot{R}_{I}^{\alpha}\dot{R}_{J}^{\beta}
\nabla_{I}^{\alpha}\nabla_{J}^{\beta}\psi_{i}^{g.s.}(\mathbf{r})
\label{eqn:gsacc}
\end{equation}
Clearly, $\ddot \psi_i^{g.s.} (\mathbf{r})$ is finite unless the ions are not moving.
At a given instant in time $\ddot \psi_i^{g.s.} (\mathbf{r})$ depends on the ionic positions $\{\mathbf{R}_{I}\}$, 
the velocities $\{\mathbf{\dot{R}}_{I}\}$, and on the orbitals $\{\delta \psi_{i}\} $. 
The dependence of $\ddot \psi_i^{g.s.} (\mathbf{r})$ on the $\{\delta \psi_{i}\}$ comes in via the 
dependence of the ionic accelerations on the $\{\delta \psi_{i}\}$ through equation \ref{eqn:ions1}.
However, equation \ref{eqn:ions1} may be rewritten as 
\begin{eqnarray}
\ddot{R}_{I}^{\alpha}  & = & 
-\frac{1}{\textrm{M}_{I}}\frac{ \partial }{\partial R_{I}^{\alpha}}\Bigg\{ E \Bigg|_{g.s.}  
+ \sum_j \int \Bigg [ \frac{\delta E} {\delta \psi_j (\mathbf{r})}\Bigg|_{g.s.} \delta \psi_j (\mathbf{r}) \nonumber \\
& + & \frac{\delta E} {\delta \psi_{j}^{*} (\mathbf{r})}\Bigg|_{g.s.} \delta \psi_{j}^{*} (\mathbf{r}) \Bigg ]
d\mathbf{r}
+ \mathcal{O}(\delta \psi^2) \Bigg\} \nonumber \\
 & = &  -\frac{1}{\textrm{M}_{I}}\nabla_{I}^{\alpha} E
\Bigg|_{g.s.} +  \mathcal{O}(\delta \psi^2) \label{eqn:ions2} 
\end{eqnarray}
from which it is clear that the dependence of $\ddot{R}_{I}^{\alpha}$, and therefore of 
$\ddot \psi_i^{g.s.}(\mathbf{r})$, on $\{\delta \psi_i \}$ is at higher than linear order. 
To first order in $\{\delta \psi_i \}$, therefore, 
equations \ref{eqn:electrons1}, \ref{eqn:taylor} and \ref{eqn:orbacc} can be combined to give
\begin{eqnarray}
\delta \ddot{\psi}_{i}(\mathbf{r})  =  -\frac{1}{\mu}\sum_j \int  \Bigg [ \frac{\delta^2 E}{\delta \psi_{j}^{*}(\mathbf{r'}) 
\delta \psi_{i}^{*}(\mathbf{r})}\Bigg|_{g.s.} \delta \psi_{j}^{*}(\mathbf{r'})  \nonumber \\
 +  \frac{\delta^2 E}{\delta \psi_{j}(\mathbf{r'}) 
\delta \psi_{i}^{*}(\mathbf{r})}\Bigg|_{g.s.} \delta \psi_{j}(\mathbf{r'})\Bigg ] d\mathbf{r'} 
- \ddot \psi_i^{g.s.}(\mathbf{r})
\end{eqnarray}
By assuming that the $\{\psi_i\}$ are either real, or are complex but with phase factors that do not
vary as the system evolves, this equation can be evaluated as
\begin{equation}
\delta \ddot{\psi}_{i}(\mathbf{r})  =  -\frac{1}{\mu}\sum_j \int 
K(i \; \mathbf{r} , j \; \mathbf{r'} ) \delta \psi_{j}(\mathbf{r'}) d\mathbf{r'}
- \ddot \psi_i^{g.s.}(\mathbf{r})  \label{eqn:electrons3}
\end{equation}
where
\begin{eqnarray}
K(i \; \mathbf{r} , j \; \mathbf{r'} ) & = & f_j \Bigg ( \delta_{ij} 
\delta(\mathbf{r}-\mathbf{r'}) H \Bigg|_{g.s.} \nonumber \\
 & + & 2\frac{\delta v(\mathbf{r})}{\delta n(\mathbf{r'})}\Bigg|_{g.s.} 
\psi^{* \; g.s.}_{j}(\mathbf{r'}) 
\psi_{i}^{g.s}(\mathbf{r})  \Bigg ) 
\end{eqnarray} 
$H$ is the Kohn-Sham single particle Hamilitonian, 
$v(\mathbf{r}) = v_{\text{H}}(\mathbf{r}) + v_{\text{XC}}(\mathbf{r})$
is the sum of the Hartree and exchange-correlation potentials, and $f_j$ is the 
occupation number of the $j^{th}$ orbital. It is clear that $K(i \; \mathbf{r} , j \; \mathbf{r'} )$ 
is independent of the $\{\delta \psi_i \}$ and therefore evolves on ionic timescales.

If the $\{ \delta \psi_i \}$ are sufficiently small, their motion is governed
by equation \ref{eqn:electrons3} which describes driven coupled oscillations of the
$\{\delta \psi_i\}$, where the driving term is given by the acceleration of the ground state $\{\ddot \psi_i^{g.s.}\}$.
By means of a suitable unitary transformation, 
$ K(i \; \mathbf{r} , j \; \mathbf{r'} )$ can be diagonalized and equation \ref{eqn:electrons3}
recast into a simpler form involving transformations of the functions 
$\{\delta \psi_{i}(\mathbf{r})\}$ and $\{\ddot \psi_i^{g.s.}(\mathbf{r})\}$, however, for the purposes
of the present discussion it is sufficient to notice that, by reducing the value 
of the free parameter $\mu$, one can make the magnitude of $K/\mu$ arbitrarily large thereby
increasing the frequencies of the oscillations to ``ultra high'' values, while the driving 
term $\ddot \psi_i^{g.s.}(\mathbf{r})$ remains constant. It is assumed here that a value of 
$\mu$ has been chosen such that there exist three distinct 
time scales $\tau_{S}\ll\tau_{I}\ll\tau_{L}$ 
in the Car-Parrinello dynamics. $\tau_{S}$ is a short time scale that is 
comparable to the period of the ultra high frequency oscillation of the orbitals, 
$\tau_{L}$ is the long time scale on which the ions move, and $\tau_{I}$ is 
an intermediate time scale. When averaged over a time scale of order $\sim \tau_{I}$, 
equation ~\ref{eqn:electrons3} becomes

\begin{equation}
\overline{\delta \ddot{\psi}_{i}(\mathbf{r})}   =   -\frac{1}{\mu}\sum_j \int 
K(i \; \mathbf{r} , j \; \mathbf{r'} ) \overline{\delta \psi_{j}(\mathbf{r'})} d\mathbf{r'}
- \ddot \psi_i^{g.s.}(\mathbf{r})  \label{eqn:electrons4}
\end{equation}
where the averaging only significantly affects $\delta \ddot{\psi}_{i}(\mathbf{r})$ 
and $\delta \psi_{i}(\mathbf{r})$ because both $K$ and $\ddot \psi_i^{g.s.}(\mathbf{r})$ 
vary on the much longer time scale $\tau_{L}$. If the orbitals are to remain 
close to the ground state, then  $\overline{\delta \ddot{\psi}_{i}} \approx 0, \; \forall i $,
because the short time scale dynamics of the orbitals should be oscillatory rather than 
accelerating systematically in one direction.
By inverting equation ~\ref{eqn:electrons4}, the average deviation of the wavefunctions
from their ground state values can therefore be written as
\begin{eqnarray}
\overline{\delta \psi_{j}(\mathbf{r'})} = -\mu\sum_i \int 
K^{-1}( j \; {\mathbf r'} , i \; {\mathbf r}) \ddot \psi_i^{g.s.}(\mathbf{r}) d\mathbf{r} \label{eqn:electrons5}
\end{eqnarray}
$\overline{\delta \psi_{j}(\mathbf{r'})}$ is, in general, non-zero and dependent only on ionic positions
and velocities. Therefore, the electronic orbitals do not average to their ground state
values on time scales much longer than the time scale of their ultra high frequency oscillations but shorter
than the time scale of ionic motion.
On the other hand, $\overline{\delta \psi_{j}(\mathbf{r'})}$ depends linearly on the fictitious mass $\mu$ and therefore
can be made arbitrarily small by reducing its value. Furthermore, as pointed out by Pastore {\em et al.}\cite{pastore}, 
by decreasing the value of $\mu$ the frequencies of the oscillations of the $\delta \psi_i$ about their slowly evolving
average values can be made so large that virtually no 
energy is transferred to them from the lower (ionic) frequency modes of the 
orbitals and ions. If the amplitudes of these oscillations are initially small, therefore, they will remain
small unless the system is artificially perturbed (by, for example, discontinuously changing the velocities
of the ions).  

To summarize this discussion of the orbital dynamics: it has just been shown that for small
values of the fictitious mass, $\mu$, the dynamics of the Car-Parrinello orbitals can be 
described as consisting of ultra high frequency oscillations about average values that evolve
on ionic timescales and that are given by equation \ref{eqn:electrons5}.  For sufficiently small
values of $\mu$ ( and $\{\delta \psi_i\}$) the deviations of the forces on the ions from
their values at the electronic ground state can therefore be separated into an ultra high frequency 
oscillatory part that averages to zero on the time scale relevant to ion dynamics 
and a systematic part that does not.  The systematic part of the errors in the forces 
on the ions is the main focus of the present work.

For larger values of $\mu$, or if the particulars of a simulation are such that 
ultra high frequency oscillations of the $\{\delta \psi_i\}$  have a large amplitude, 
contributions to the errors in the forces that are of higher than
linear order may play a role, even if the separation of time scales $\tau_{S}\ll\tau_{L}$
can still be maintained, however, such effects will not be discussed in the present work.

\subsection{The forces on the ions} \label{section:ions}
In this section, the impact of the orbital dynamics discussed in the
previous section on the forces on the ions are discussed. Of primary concern 
is the systematic part of the error in the forces due to the displacement
of the average values of the orbitals from the ground state, $\overline{\delta \psi}$. This contribution
is now derived.

Equation \ref{eqn:ions1} can be written as
\begin{eqnarray}
 F_{{I}}^{\alpha} & = & 
- \frac{\partial E}{\partial R_{I}^{\alpha}} \nonumber \\
& = &
- \nabla_{I}^{\alpha}E 
+\sum_{i}\int \Bigg [\frac{\delta E}{\delta \psi^{*}_{i}(\mathbf{r})}
\frac{\partial \psi^{*}_{i}(\mathbf{r})}{\partial R_{I}^{\alpha}} + \frac{\delta E}{\delta \psi_{i}(\mathbf{r})}
\frac{\partial \psi_{i}(\mathbf{r})}{\partial R_{I}^{\alpha}} \Bigg ] d\mathbf{r} \nonumber \\
\label{eqn:fcp1}
\end{eqnarray}
Substitution of equation~\ref{eqn:electrons1} yields
\begin{equation}
F_{{I}}^{\alpha}   = 
-\nabla_{I}^{\alpha}E -
\sum_{i}\mu \int \Bigg [ \ddot{\psi}_{i}(\mathbf{r})
\nabla_{I}^{\alpha} \psi^{*}_{i}(\mathbf{r})
+ \ddot{\psi}^{*}_{i}(\mathbf{r})
\nabla_{I}^{\alpha} \psi_{i}(\mathbf{r})\Bigg ] d\mathbf{r}
\label{eqn:fcp2}
\end{equation}
since $\nabla_{I}^{\alpha} \psi_{i}(\mathbf{r}) = \partial \psi_{i}(\mathbf{r})/\partial R_{I}^{\alpha}$.
The first term on the right-hand side of equation~\ref{eqn:fcp2} is now expanded
in a Taylor series about the ground state :
\begin{eqnarray}
-\nabla_{I}^{\alpha}  E  & = & 
-\nabla_{I}^{\alpha}\bigg\{E
\Bigg |_{g.s.}  + \sum_{i}
\int \Bigg [
\frac{\delta E}{\delta \psi^{*}_{i}(\mathbf{r})}\Bigg |_{g.s.}
\delta \psi^{*}_{i}(\mathbf{r}) \nonumber \\
 & + & \frac{\delta E}{\delta \psi_{i}(\mathbf{r})}\Bigg 
|_{g.s.}\delta \psi_{i}(\mathbf{r})\Bigg ] d\mathbf{r}
 + \mathcal{O}(\delta \psi_{i}^{2})
\bigg \} \nonumber \\
 & = &  F_{BO_{I}}^{\alpha} + 0 + \mathcal{O}(\delta \psi_{i}^{2}) 
\label{eqn:fcp3}
\end{eqnarray}
where $F_{BO_{I}}^{\alpha}$ is the $\alpha$-th Cartesian component of the Born-Oppenheimer force 
 on atom $I$.
As pointed out in section ~\ref{section:orbitals}, on time scales relevant to ionic motion $\overline{\ddot{\psi}_i(\mathbf{r})} 
= \overline{\delta \ddot{\psi}_i (\mathbf{r})} + \ddot \psi_i^{g.s.} (\mathbf{r}) = \ddot{\psi}_i^{g.s.}(\mathbf{r})$,
therefore, equations \ref{eqn:gsacc}, \ref{eqn:fcp2}, and \ref{eqn:fcp3} can be combined to give 
the error in the Car-Parrinello force on timescales 
( $ > \tau_{I}$) relevant for ionic motion as\cite{error_footnote}
\begin{eqnarray}
\Delta F_{I}^{\alpha}&=&\overline{F_{{I}}^{\alpha}} - F_{BO_{I}}^{\alpha} 
\nonumber\\ &=& -2\mu \sum_{i}\Re\Bigg\{ \int \bigg [ \sum_{J, \beta} 
\ddot{R}_{J}^{\beta} 
\nabla_{I}^{\alpha} \psi_{i}^{* \; g.s.} (\mathbf{r}) \nabla_{J}^{\beta}\psi_{i}^{g.s.}
(\mathbf{r})  \nonumber \\
  &+&   \sum_{J,\beta,K,\gamma} \dot{R}_{J}^{\beta}\dot{R}_{K}^{\gamma}
\nabla_{I}^{\alpha}\psi_{i}^{*\;g.s.}(\mathbf{r}) \nabla_{K}^{\gamma}
\nabla_{J}^{\beta}\psi_{i}^{g.s.}(\mathbf{r})\bigg ]\text{d}\mathbf{r} \Bigg\}\nonumber\\
\label{eqn:error}
\end{eqnarray}
This expression is valid in the limit of small $\{ \delta \psi_i \}$ and depends
only on ionic positions and velocities. It has no first order dependence on the $\{\delta \psi_i\}$. 

This systematic error in the Car-Parrinello forces arises from the fact that the electronic
ground state evolves on ionic time scales. As ions move, they must push the orbitals around.
The orbitals carry inertia and so the ions must impart momentum to them in order for them
to follow the ground state. This means that the ions lose momentum to the orbitals, i.e. the
ions encounter a resistance to their motion. It is as though the ions move through a viscous, 
inhomogeneous and time-varying medium.

The right hand side of equation \ref{eqn:error} depends on the
details of the electronic ground state, as well as the velocities of the ions, and so it 
is difficult to make general statements about how $\Delta F$ affects the dynamics
and thermodynamics of a simulation.
In the absence of further information there
is no reason to expect either dynamical properties or
thermodynamic properties to be correct if $\mu$ is
large enough to make the error of equation \ref{eqn:error} significant.
However, in the limit in which individual ions are spherically-symmetric and
do not interact with one another, this error has been shown to reduce to the form 
$\Delta F_{I}^{\alpha} = -\Delta \textrm{M} \ddot{R}^{\alpha}_{I}$, where $\Delta \textrm{M}_{I}$ is
a constant that is proportional to the total 
{\em quantum} kinetic energy of the electronic states on ion $I$ \cite{us,blochl2}. 
In other words, if ions don't interact with one another (i.e. in the limit of the infinitely
dilute gas\cite{blochl2}) the systematic error in the force due to the displacement 
of the average values of the orbitals from the ground state, 
$\{\overline{\delta \psi_i} \}$, has the same effect as simply increasing
the masses of the ions by an amount $\Delta \textrm{M}_{I}$. 
This result had been noticed much earlier\cite{blochl_and_parrinello}, and
described in reference \onlinecite{blochl_paw}. The more general result of
reference \onlinecite{us} was also independently found by Bl\"ochl\cite{blochl2}.
An equation of motion for the ions that is an approximation to the equation of motion
of the ions in CPMD is therefore given by\cite{us,error_footnote}
\begin{eqnarray}
(\textrm{M}_{I}+\Delta\textrm{M}_{I})\ddot{R}_{I}^{\alpha} = 
-\frac{ \partial E [\{\psi_{i}^{g.s.}\},\{\mathbf{R}_{I}\}]} 
{\partial R_{I}^{\alpha}} = F_{BO_{I}}^{\alpha}
\label{eqn:ions3}
\end{eqnarray}
The degree to which equation \ref{eqn:ions3} describes the motion of the ions in CPMD
depends on the degree to which the electronic structure of the condensed system 
in question can be described in terms of a linear superposition of tightly bound
ionic orbitals that are transported rigidly (i.e. without changing shape in response to
their environment) as the ions move. This approximation to CPMD is known as the
``rigid-ion'' approximation\cite{us}.
Equation \ref{eqn:ions3} can be understood as arising
from the fact that such tightly bound orbitals have inertia ($\propto \mu$) and 
that this adds to the total inertia of the ions: the ions have to 
carry the ``heavy'' orbitals.

In a dynamics described by equation \ref{eqn:ions3}, i.e. a Born-Oppenheimer dynamics but with
increased ionic masses, thermodynamic
properties are unchanged. This is only the case, however, if thermodynamic
quantities such as temperature and thermal pressure are calculated
using these increased values of the masses. In reference
\onlinecite{us} this was demonstrated for the case of MgO: The true temperature
of the oxygen subsystem was that calculated using a mass for the oxygen ions
of $\textrm{M}_{\text{O}}+\Delta\textrm{M}_{\text{O}}$, where $\textrm{M}_{\text{O}}$ was the 
bare oxygen mass of $16$ a.m.u. and $\Delta\textrm{M}_{\text{O}}$ a correction that
accounted for the inertia of the orbitals moving rigidly with the oxygen ions.  
Car-Parrinello simulations that neglect the correction to the temperature due to 
increased ionic masses have effectively been performed at temperatures that 
are higher than those reported. Furthermore, this correction is important if 
more than one ionic thermostat is used in a simulation. For example, if mass 
corrections are not accounted for in the calculation of temperature, and 
if a different thermostat is applied to each ionic species, the result can
 be that each species is effectively maintained at a different temperature.
The definition of temperature may be particularly important if so-called 
``massive thermostating'' is employed\cite{massive}, in which each degree 
of freedom is coupled to its own thermostat.

In a system in which there is only one atomic species and every atom is in
an identical chemical environment and if the rigid-ion approximation works well, 
dynamical properties can be corrected simply by rescaling time by a factor $\sqrt{M/(M+\Delta M)}$.
If the rigid-ion approximation works, but there is more than one atomic species, 
dynamical properties can only be made correct by rescaling the mass of each
atom {\em a priori}. In other words, the simulation is performed with a nominal 
mass for each ion $I$ of $M_{I} -\Delta M_{I}$ so that the mass of 
the ion once 'dressed' by the heavy Car-Parrinello orbitals is the 
correct mass $M_{I}$. If this is not done, large errors in dynamical properties can result\cite{us}.

Strictly speaking, the rigid ion approximation is not applicable when ions
interact with one another. In other words, when there exist electronic orbitals $\phi$ 
that depend on the positions of more than one ion, i.e.
for which $\mathbf{\nabla}_{I}\phi \ne  0$ and 
$\mathbf{\nabla}_{J}\phi \ne 0$
for some ions $I\ne J$. 
Of course, this is generally the case for all condensed matter
and therefore for all systems of interest. What this means is that 
the dynamics and thermodynamics of CPMD always differ from BOMD.
Fortunately, for all condensed systems tested to date, most of the error 
of equation \ref{eqn:error} can be corrected by applying mass corrections
to the ions. 
The concern that remains, however, is that even though it is
much smaller, the part of the error that remains once 
mass corrections have been applied may not be negligible.

A further difficulty associated with correcting the masses of the ions
is that the quantum kinetic energy of the electronic states on a given 
atomic species is not a well defined quantity. 
This is because each orbital cannot, in general, be associated with only one atom.
A method must therefore be devised to approximate the mass corrections. 
In the simulations of D$_2$O presented here, the mass corrections that are 
used are those that minimize the errors in the forces on the ions. 

In order to differentiate between the part of $\Delta \mathbf{F}$ that amounts to 
a simple mass correction, and the part that remains once this mass correction
has been applied, the quantities $\Delta \mathbf{F}^{m}$ and $\Delta \mathbf{F}^{r}$ are introduced
where $\Delta \mathbf{F}^{m} = -\Delta \textrm{M} \ddot{\mathbf{R}}$ 
and $\Delta \mathbf{F}^{r} = \Delta \mathbf{F} - \Delta \mathbf{F}^{m}$
is the remaining error.
The expression for $\Delta \mathbf{F}$ in equation \ref{eqn:error} has been derived by taking an average over 
the ultra high frequency orbital oscillations, and 
therefore a more complete expression for the Car-Parrinello force is 
\begin{eqnarray}
\mathbf{F}_{I} & = & \mathbf{F}_{BO_{I}}(\{\mathbf{R}_{I}\}) 
+ \Delta \mathbf{F}_{I}(\{\mathbf{R}_{I}\},\{\mathbf{\dot{R}}_{I}\}) \nonumber \\
& + & \delta \mathbf{F}_{I}(\{\delta \psi_i\}) + \mathcal{O}(\delta \psi^2) \nonumber \\
 & \approx & \mathbf{F}_{BO_{I}} 
+ \Delta \mathbf{F}_{I}^{m} + \Delta \mathbf{F}_{I}^{r}
+ \delta \mathbf{F}_{I}
\end{eqnarray}
where $\delta \mathbf{F}_{I}$ is the oscillatory part of the
force which, for small enough $\mu$, averages to zero on time scales $\tau_I \gg \tau_S$.

\begin{figure} 
\epsfig{figure=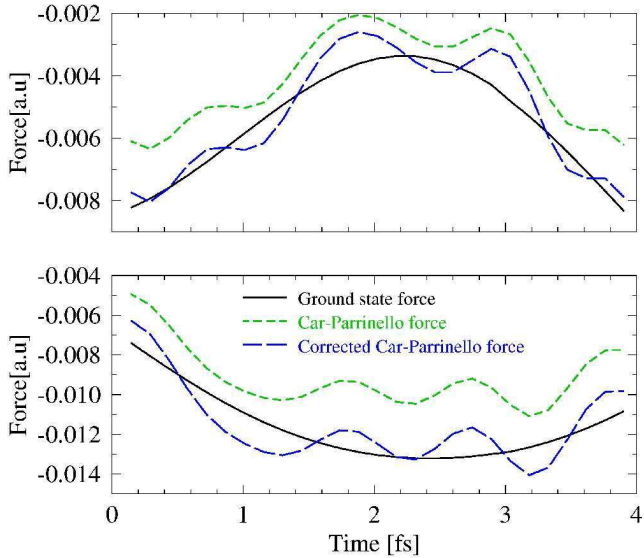,width=8.5cm}
\caption{Some samples of force components on oxygen ions along a $4$ fs segment of 
CPMD trajectory $1$ ps after temperature has been adjusted using velocity rescaling. 
The Car-Parrinello force and the Car-Parrinello force
once corrected with the best constant mass correction is compared to the
ground state (Born-Oppenheimer) force, $F_{BO_{I}}$
at the same ionic positions.}
\label{fig:jumpy}
\end{figure}

Figure \ref{fig:jumpy} demonstrates that, indeed, the difference between the Born-Oppenheimer
and Car-Parrinello forces consists of both a systematic
part $\Delta \mathbf{F}$, and an oscillatory part $\delta \mathbf{F}$. The systematic part 
appears to be mostly corrected by the application of mass corrections to the ions. 
The forces in figure \ref{fig:jumpy} are those from a Car-Parrinello simulation in which 
velocity rescaling was employed for $0.5$ ps to bring the temperature from $T \sim 120 K$ to 
$T\sim 220$ K. The velocities of the ions were adjusted only $4$ times, in total, during this $0.5$ ps . 
The system was then equilibrated for one picosecond after which the forces were examined along the 
$4$ fs segment of trajectory and compared to the forces at the same ionic positions, but with the
orbitals in their ground state, i.e. the Born-Oppenheimer forces $\mathbf{F}_{BO_{I}}$.
The amplitude of the oscillations, $\delta \mathbf{F}$, in figure \ref{fig:jumpy} appears 
worryingly large. It will be shown in sections \ref{section:details} and \ref{section:forces}, however,  
that if care is taken to ensure that the velocities of the ions are not changed
discontinuously, the large amplitude of these oscillations disappears and the 
contribution of $\delta \mathbf{F}$ to the instantaneous error in the Car-Parrinello force can be neglected.
At higher temperatures, however, large amplitude oscillations may always be
present, as was the case in simulations of MgO in reference \onlinecite{us}.

In section \ref{section:forces}, the systematic part of the 
error $\Delta \mathbf{F}$, will be examined. It has been assumed in the past\cite{kuo} that
$\Delta \mathbf{F} \approx \Delta \mathbf{F}^{m}$ and that $\Delta \mathbf{F}^{r}$
can be neglected.  Therefore, in section \ref{section:forces}, the validity of this assumption
will be investigated.
The degree to which $\Delta \mathbf{F}$ can be eliminated by applying mass corrections
to the ions will be tested. In other words, the magnitude of $\Delta \mathbf{F}^{r}$,  
which has the potential to alter thermodynamic properties and for which no correction 
yet exists, will be examined.

\section{Simulation Details} \label{section:details}
\begin{figure*} 
\epsfig{figure=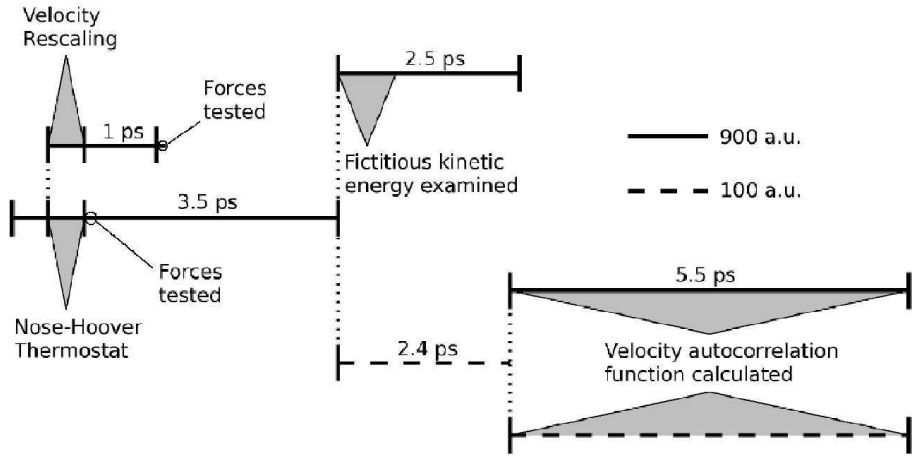,width=12.0cm}
\caption{A schematic illustrating the sequence of Car-Parrinello molecular dynamics simulations that
were performed.}
\label{fig:sequence}
\end{figure*}
In the CPMD simulations of D$_2$O reported here norm-conserving pseudopotentials
have been used to describe oxygen\cite{trouillier_martins} and 
deuterium\cite{gygi} and the valence electronic
orbitals were expanded in plane waves with a maximum energy of 70 Rydberg.
Simulations were performed on a $24\times 24 \times 12 $ a.u. simulation box containing
$32$ D$_2$O molecules. Only the $\Gamma-$point was used to sample the Brillouin zone.
A gradient-corrected functional was used to treat the effects of exchange and correlation\cite{blyp}.
In a preliminary CPMD simulation of liquid water mass corrections of 
$\Delta M_{O} = 6.77 \mu$ a.u. and $\Delta M_{D} = 0.213 \mu$ a.u. ,
were found for the oxygen and deuterium ions, respectively. 
These mass corrections were found
by calculating the ground state forces at selected points along a trajectory and by
minimizing the error in the CPMD forces relative to these ground state forces. 
These mass corrections were then used in all of the CPMD simulations 
of ice reported here by reducing the values of the ionic 
masses used in calculating the acceleration
from Newton's equation of motion i.e.
\begin{equation}
\ddot{R}_{I}^{\alpha} = -\frac{1}{\textrm{M}_{I}-\Delta \textrm{M}_{I}}
\frac{ \partial E [\{\psi_{i}\},\{\mathbf{R}_{I}\}]}
{\partial R_{I}^{\alpha}} 
\label{eqn:ions4}
\end{equation}
The values of the fictitious mass used in the simulations
reported here were $900$ a.u. and, for comparison, $100$ a.u.  
The equations of motion for electrons and ions were integrated using 
a molecular dynamics time step of $6$ a.u. 
for the $\mu = 900$ a.u. simulations and $2$ a.u. for the $\mu = 100$ 
a.u. simulations. No mass preconditioning 
scheme was used in any of the simulations reported here.

Much care has been taken to minimize errors in the simulations reported here. 
For example, as pointed out by Remler and Madden\cite{remler}, and discussed
in section \ref{section:ions} it is important that the velocities
of the ions are not changed discontinuously as this can give a sudden kick to 
the electronic orbitals that results in large-amplitude ultra high frequency oscillations of 
the orbitals around their equilibria. 
Great care has therefore been taken to eliminate
this source of error from the simulations reported here with the result (evident in 
figures \ref{fig:force1} and \ref{fig:force2} of section \ref{section:forces}) that
instantaneous errors in the forces due to ultra high frequency oscillations
of the orbitals (i.e. $\delta \mathbf{F}$ ) are small enough relative to the 
systematic errors of equation \ref{eqn:error} that they may safely be neglected.

\begin{figure} 
\epsfig{figure=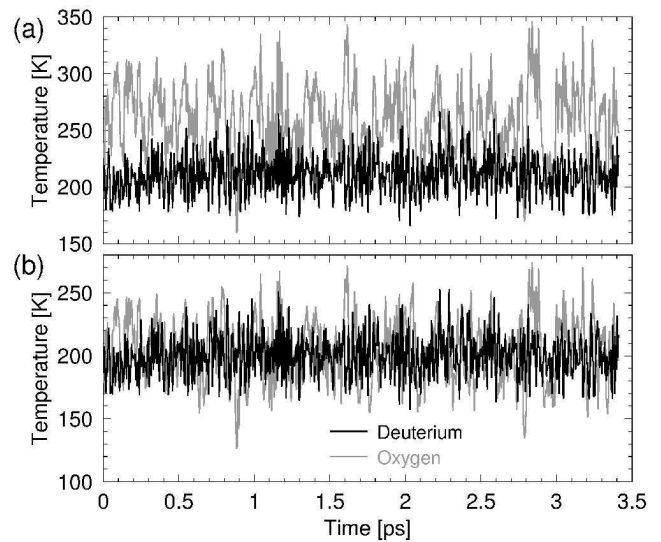,width=8.5cm}
\caption{The temperatures of the oxygen and deuterium subsystems during the $3.5$ ps equilibration
simulation with $\mu = 900$ a.u. In (a) the temperatures have been calculated using
masses for the ions that have been corrected to account for the extra inertia due to the
weight of the Car-Parrinello orbitals (in this case $16$ a.m.u and $2$ a.m.u. due to 
the {\em a priori} rescaling of the masses). The oxygen and deuterium subsystems appear to be 
out of thermal equilibrium. In (b) the temperatures have been calculated
in the standard way using the bare ionic masses ($12.659$ a.m.u and $1.895$ a.m.u). 
The oxygen and deuterium subsystems appear perfectly equilibrated.}
\label{fig:temperature1}
\end{figure}

The input coordinates for the CPMD were obtained by performing a very 
long molecular dynamics simulation of ice at low temperature
($\sim 100$ K ) using an {\em ab initio} parametrized polarizable atomistic potential
of the same form as that constructed for silica in reference~\onlinecite{us_silica}.
This potential does not provide a very realistic description of water but it
was deemed preferable to using randomized initial coordinates. 
From these initial conditions a sequence of CPMD simulations were performed
as shown schematically in figure~\ref{fig:sequence}.
CPMD simulations were begun with the electrons in their ground state and the
ions at zero velocity. A fictitious mass of $\mu = 900$ a.u was used and therefore the
ionic masses of oxygen and deuterium were rescaled to $M_{O} - \Delta M_{O} = 12.659$ a.m.u 
and $M_{D} - \Delta M_{D} = 1.895$ a.m.u, respectively and temperature was calculated using the true
ionic masses of $M_{O} = 16$ a.m.u and $M_{D} = 2$ a.m.u.
After half a picosecond of simulation in the microcanonical ensemble, 
the ions were at a temperature of $120$ K. At this point two separate
continuations of the simulation were performed.
The first continuation was performed in order
to demonstrate the different contributions to the error in the forces on the ions and 
the effect of velocity rescaling on these forces, and
has been discussed in section \ref{section:ions}.
In the second continuation, a Nos\'e-Hoover thermostat\cite{nose} was attached to the ions and the temperature
was smoothly increased to approximately $250$ K during a further half picosecond of simulation. 
The system was then equilibrated without a thermostat for $3.5$ picoseconds. 
As shown in figure~\ref{fig:temperature1} the temperatures of the subsystems of oxygen and deuterium
ions were calculated and compared with and without the use of the corrections to the masses of
the ions in the definition of temperature. As can be seen, when mass corrections are used the
oxygen and deuterium subsystems are at different temperatures indicating that the system
is not well equilibrated. 
If mass corrections are not used, the system appears perfectly equilibrated. 
At this point, the simulation was again continued in two separate simulations. 
In one of these simulations the value of $\mu = 900$ a.u. was used as before. 
In another simulation the fictitious mass was changed to $\mu = 100$ a.u. 
The rescaled masses of the ions in the simulation 
with $\mu = 100$ a.u. were $15.629$ a.m.u. and $1.988$ a.m.u. 
If the rigid-ion approximation was perfectly applicable the 
fact that the ionic masses have been rescaled {\em a priori} would mean that 
the two simulations with different values of $\mu$ should have identical dynamic and
thermodynamic behaviour.

\begin{figure} 
\epsfig{figure=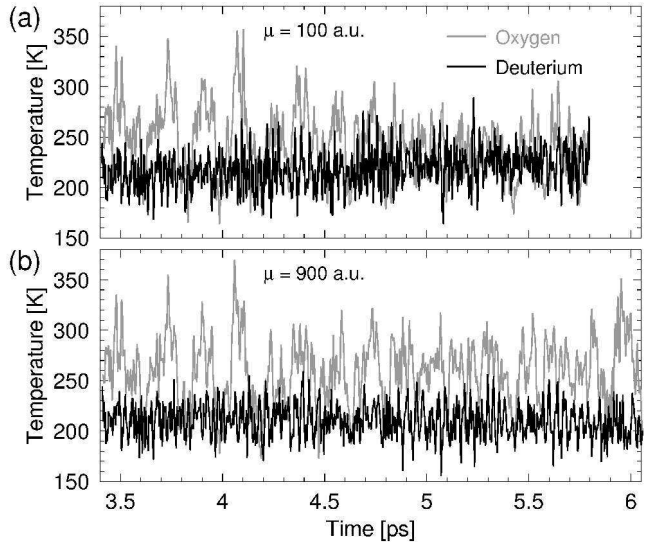,width=8.5cm}
\caption{The temperatures of the oxygen and deuterium subsystems calculated using the 
corrected ionic masses ($M + \Delta M$)
in the two continuations of the simulation shown in figure~\ref{fig:temperature1}. In (a)
a fictitious mass of $100$ a.u. has been used and thermal equilibration between the subsystems is
observed. In (b) a fictitious mass of $900$ a.u. is used and the subsystems 
remain out of thermal equilibrium.}
\label{fig:temperature2}
\end{figure}
It was found (figure \ref{fig:temperature2}) that after more than a further $2.5$ ps 
the simulation with $\mu=900$ a.u. still showed no sign of thermalization according 
to the mass-corrected definition of temperature.
However, the simulation with $\mu=100$ a.u. equilibrated quickly and the subsystems
were at the same temperature after $1.5 - 2$ ps (figure ~\ref{fig:temperature2}). 
In order to compare the phonon spectra of ice using these two 
different values of $\mu$ 
for reasonably well equilibrated simulations at the same temperature 
it was decided to continue {\em both} simulations from the end of this initial $2.4$ ps run
with $\mu = 100$ a.u. For both $\mu = 100$ a.u. and $\mu = 900$ a.u., a further $5.5$ ps of simulation
were carried out during which time the oxygen and deuterium subsystems remained at the same
(mass-corrected) temperature in both simulations. 
For both values of $\mu$, the velocity autocorrelation function was calculated on a 
time domain of $1.45$ ps by averaging over this final $5.5$ ps of simulation.
These velocity autocorrelation functions were then fourier transformed to obtain 
the phonon power spectra.

The reason for the inability of the $\mu = 900$ a.u. simulations to thermalize 
during the first $7$ ps of simulation remains unclear. 
By using the dressed ionic masses in the calculation of temperature we are implicitly 
assuming that orbitals move rigidly with the ions and are unperturbed by their environment.
In a condensed system this is an approximation and deviations from
the rigid-ion limit always occur. In the opposite limit to the rigid-ion
approximation, i.e. in the limit of very weak orbital-ion coupling,
the temperature of the ions should be calculated with the bare ionic masses.
For systems that are not perfectly ionic, therefore, the correct
definition of temperature is unclear. While the results of section \ref{section:forces} suggest
that a constant mass correction for the ions is not perfectly appropriate for D$_2$O, the 
results of section \ref{section:timescale} suggest that the rigid-ion approximation does
a remarkably good job of estimating the fictitious kinetic energy of the orbitals.
In addition, the fact that during the final $5.5$ ps of simulation the oxygen and deuterium 
subsystems remained at the same mass-corrected temperature suggests that it is appropriate
to calculated temperature using rescaled masses. 

The fact that reducing the value of $\mu$ appeared to facilitate thermalization may indicate
that the thermalization problem was an artifact of the fictitious mass $\mu$, however, 
further work is required to clarify this issue.

\section{The time scale of orbital motion} \label{section:timescale}
A common view of CPMD is that, if a large enough gap exists between the energies of the 
occupied and unoccupied electronic states, the electronic orbitals move on time scales 
that are much faster than typical ionic time scales. For example, it has been suggested
that the motion of the orbitals in CPMD can be approximately described as a superposition
of oscillations whose frequencies are given by 
$\omega_{ij} = ( \frac{2 (\epsilon_j - \epsilon_i)}{\mu})^{1/2}$, where $\epsilon_i$ and 
$\epsilon_j$ are the Kohn-Sham energy eigenvalues of 
occupied and unoccupied electronic states, respectively\cite{marx,grossman,caryip}.
In a system with a substantial energy gap between occupied and unoccupied states, all these
frequencies are very high compared to typical ionic frequencies. In this view of CPMD the
lowest vibrational frequency present in the orbital subsystem is determined by the size of the
energy gap and the fictitious mass.
However, it should be clear from the work of Pastore {\em et al.} and sections 
~\ref{section:background} and ~\ref{section:theory} of the present work that, 
if the orbitals are to remain close to the ground state and the ions are moving, 
their motion must contain an ionic time scale component. 
Here it is shown that, in fact, it 
makes the dominant contribution to the total orbital kinetic energy. 
The slow orbital motion results from the coupling between ions and the orbitals
and it is precisely this coupling that leads to a systematic departure of the average 
forces in CPMD from the ground state forces.

If the rigid ion approximation holds perfectly, then the part of the kinetic energy 
of the orbitals due to the evolution of the electronic ground state can be obtained simply 
from the mass corrections and the velocities of the ions, i.e.
\begin{equation}
\sum_{i}\mu \langle \dot{\psi}_{i} | \dot{\psi}_i \rangle = 
\frac{1}{2}\sum_{I\epsilon O}\Delta M_{O} \dot{R}_{I}^{\alpha}\dot{R}_{I}^{\alpha} 
+ \frac{1}{2}\sum_{I\epsilon D}\Delta M_{D} \dot{R}_{I}^{\alpha}\dot{R}_{I}^{\alpha}
\label{eqn:fke}
\end{equation}
\begin{figure} 
\epsfig{figure=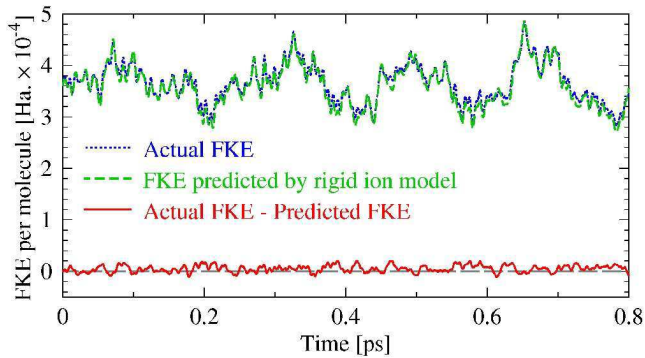,width=8.5cm} 
\caption{The fictitious kinetic energy (FKE) of the electronic orbitals as a 
function of time compared to the FKE predicted by assuming that 
orbitals are inert and rigidly follow the ions. 
The close agreement between the curves verifies that orbital 
motion occurs primarily on ionic time scales and
that the contribution of the ultra-high frequency oscillations of 
the orbitals about their instantaneous equilibria to the total FKE is tiny by comparison.}
\label{fig:fke}
\end{figure}
In figure ~\ref{fig:fke} the fictitious kinetic energy (FKE) as predicted
by the rigid ion model in equation ~\ref{eqn:fke} is compared to the FKE that
is extracted from the CPMD simulation. 
The difference between them is also plotted and this is made up both of ultra high 
frequency oscillations of the orbitals about their instantaneous equilibria and the 
part of the ionic time scale evolution of the ground state that cannot be described 
within the rigid ion approximation.
The very close agreement between the predicted FKE and the actual FKE
demonstrates clearly that electronic orbitals move on 
ionic time scales and, furthermore, that this slow motion accounts for almost all of their
fictitious kinetic energy. This also demonstrates that the lowest vibrational
frequencies of the orbital subsystem are almost\cite{almost} independent of $\mu$ and are simply
given by the lowest vibrational frequencies of the ionic subsystem.
Some previous simulations have calculated the orbital
vibrational power spectrum by fourier transforming the orbital velocity autocorrelation
function and have not detected such low frequencies\cite{pastore,grossman}.
However, in these simulations the orbital vibrations were analysed while the
ions were kept stationary and therefore the contribution to the orbital motion
from the evolution of the electronic ground state was not present.
In reference \onlinecite{herbert} Herbert and Head-Gordon have analysed the orbital vibrations 
during ion dynamics, and figure 7 of that paper clearly shows a contribution to the orbital
frequency spectrum that has a low (ionic) frequency and that appears almost independent of the 
fictitious mass.

\section{Forces} \label{section:forces}
The goal of first principles molecular dynamics is to make a direct connection between
an accurate description of the electrons and an accurate calculation of the physical
property of interest.
It is important to know that a calculated physical property that agrees well
with experiment does so because the trajectory is realistic due to good forces
calculated from a good treatment of the electronic structure. When there is
a microscopic basis for agreement with experiment
the method can be applied with some confidence to situations or to properties for which 
experimental data is unavailable. 
This point is stressed because agreement with experiment in molecular dynamics
simulations can often occur due to a cancellation of errors. This has been 
observed both in first principles simulations\cite{grossman} and in simulations using
empirical potentials where potentials that agree with experiment on many properties\cite{bks}
have been found to provide a very poor microscopic description of the interatomic
forces\cite{us_silica}.

It is obvious, therefore, that in order to proceed without recourse to
empiricism or {\em a posteriori} experimental verification one should
be able to depend on the accuracy of the computed forces.
What is much less obvious is how one should judge the accuracy of forces. 
One way of quantifying the average magnitude of the departure of a set of forces $\{F_{I}\}$ 
from some reference forces $\{F^{\text{ref}}_{I}\}$ is by computing the percentage
root-mean-squared difference between the two sets of forces relative to the root-mean-squared 
force component, i.e.  : 
\begin{equation}
\Delta_{rms} = 100\times 
\frac{\sqrt{\sum_{I,\nu}\|F_{I}-F^{\text{ref}}_{I}\|^{2}}}
{\sqrt{\sum_{I,\nu}\|F^{\text{ref}}_{I}\|^{2}}} 
\label{eqn:rms}
\end{equation}
where the sum $\sum_{I,\nu}$ is over a large number of ions $I$ and
over a large number of points $\nu$ along the molecular dynamics trajectory.
In general, this is an extremely crude measure of the departure of the forces
from the reference forces, however there is frequently no alternative to using it.
A number of classical potentials have been parametrized by minimizing this quantity
while using ground state density functional theory forces as the reference. It 
has been found that, for some simple oxides, values as low as $\Delta_{rms}= 5 - 20 \%$ can 
be achieved and that, as long as $\Delta_{rms}$ is sufficiently small (i.e. $< 20\%$ ), 
the accuracy of these force fields for many experimental quantities is well correlated
with its value\cite{us_silica,us_mgo}.
Although it is crude and unsatisfactory, in the absence of an alternative quantitative 
general measure of the quality of forces in the presence of errors of unknown consequence, 
this quantity is used in the present work.

\begin{figure} 
\epsfig{figure=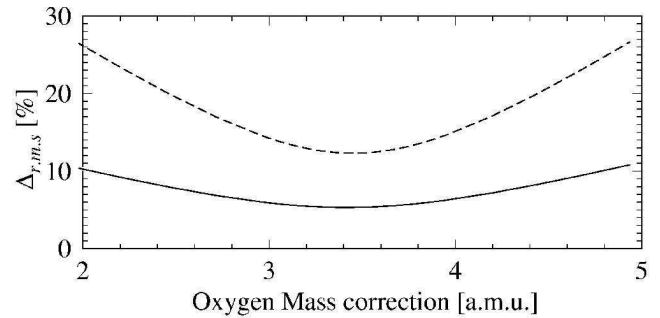,width=8.5cm} 
\caption{ The r.m.s relative error $\Delta_{r.m.s}$ in the mass-corrected forces on the oxygen ions (full line)
and on the D$_2$O molecules on the sample segment of the CPMD trajectory as a function of the 
magnitude of the mass correction. $\mu = 900$ a.u. in this simulation.}
\label{fig:correction}
\end{figure}

In this section the forces on the ions in the CPMD simulations of ice are inspected
along a segment of the trajectory.
After approximately $1$ ps of simulation the ground state forces were 
calculated along a $21$ fs segment of the CPMD trajectory. 
These ground state forces were then used as the reference forces for calculating the 
root-mean-squared (r.m.s) relative error in the CPMD forces according to equation \ref{eqn:rms} both
before and after they had been corrected using the rigid-ion mass correction. In other words (using the
notation introduced in section \ref{section:ions} and assuming that $\delta F_{I}^{\alpha}$ is negligible) 
the r.m.s. values of $\Delta F_{I}^{\alpha}$ and $\Delta F_{I}^{r \; \alpha}$ relative to the r.m.s. 
value of $F_{BO_{I}}^{\alpha}$ were computed.
Before mass corrections were applied it was found that, for oxygen ions, $\Delta_{r.m.s}$ amounted to $21.8\%$.
In other words, the r.m.s. error in the forces on the oxygen ions was $21.8\%$ of the
r.m.s. oxygen ion force component. For the deuterons $\Delta_{r.m.s}$ was $5.5 \%$.
The force on each atom is dominated by intra-molecular interactions which
are relatively constant (in the reference frame of the molecule) at room temperature and 
therefore are generally of much less interest to simulators than the inter-molecular interactions.
For this reason, the most important quantity to examine is the net force on each water molecule.
It was found that $\Delta_{r.m.s}$ for the net forces on the water molecules  was $44.7\%$.
The values of the mass corrections for the oxygen and deuterium ions were varied so as to minimize
$\Delta_{r.m.s}$ for the corrected forces.
It was found that by applying mass corrections
of $\Delta M_O = 6.91\mu = 3.41$ a.m.u and $\Delta M_D = 0.215 \mu = 0.11$ a.m.u the errors could be reduced 
substantially to $5.3 \%$ for oxygen, $1.4 \%$ for deuterium and
$12.4 \%$ for the water molecules. These mass corrections are very close to those
that were applied {\em a priori} based on preliminary tests of liquid water but in fact
it was found that, for oxygen, a wide range of values of $\Delta M_O$ 
gave very similar values for the average error.  The value of $\Delta_{r.m.s.}$ for
the corrected oxygen and water molecule forces is plotted in figure \ref{fig:correction} as
a function of the mass correction for the oxygen ion.
$\Delta_{r.m.s.}$ is found to be reasonably insensitive to the mass correction near 
the average optimal value because the optimal mass correction 
varied from atom to atom or, for a given atom, it varied in time. By changing the value of 
$\Delta M_O$ one improved the agreement of some of the forces with the
ground state forces and disimproved others. It was also found that 
the mass correction that gave best agreement between ground state and corrected CPMD forces
on a given atom was different, in general, for the different Cartesian components.
In other words, the effective masses of the ions in this simulation were time-varying tensor
quantities as is to be expected from equation \ref{eqn:error}. 

\begin{figure} 
\epsfig{figure=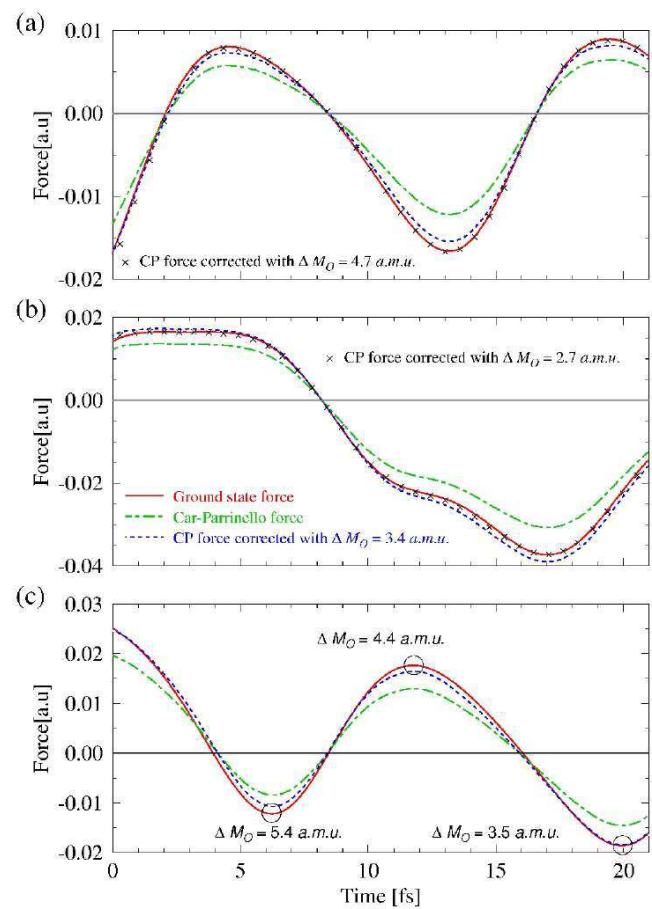,width=8.5cm} 
\caption{Some samples of force components on oxygen ions along a $21$ fs segment of 
CPMD trajectory. The Car-Parrinello force (green) and the Car-Parrinello force
once corrected with the best constant mass correction (blue) is compared to the
ground state force at the same ionic positions. Also plotted in (a) and (b) are the Car-Parrinello
forces once corrected using the optimal mass correction for that ion along this particular 
segment of trajectory. In (c) the optimal mass correction at three points along
the trajectory are indicated.}
\label{fig:force1}
\end{figure}
In figure \ref{fig:force1} some sample force components on oxygen ions are shown to illustrate
this fact. Figure \ref{fig:force1} (a) and figure \ref{fig:force1} (b) are examples of oxygen ions
which, over this segment of trajectory, have different effective masses. Fig. \ref{fig:force1} (c)
is an example of an oxygen ion force component for which the optimal mass correction varies considerably over this
short length of trajectory. The large oscillations of the Car-Parrinello forces of period $\sim 1$ fs that were
visible in figure \ref{fig:jumpy} are completely absent in figure \ref{fig:force1}, demonstrating
that the contribution, $\delta \mathbf{F}$, of the ultra high frequency oscillations of the orbitals to the force
is negligible due to the ions having been accelerated continuously throughout this simulation.

\begin{figure} 
\epsfig{figure=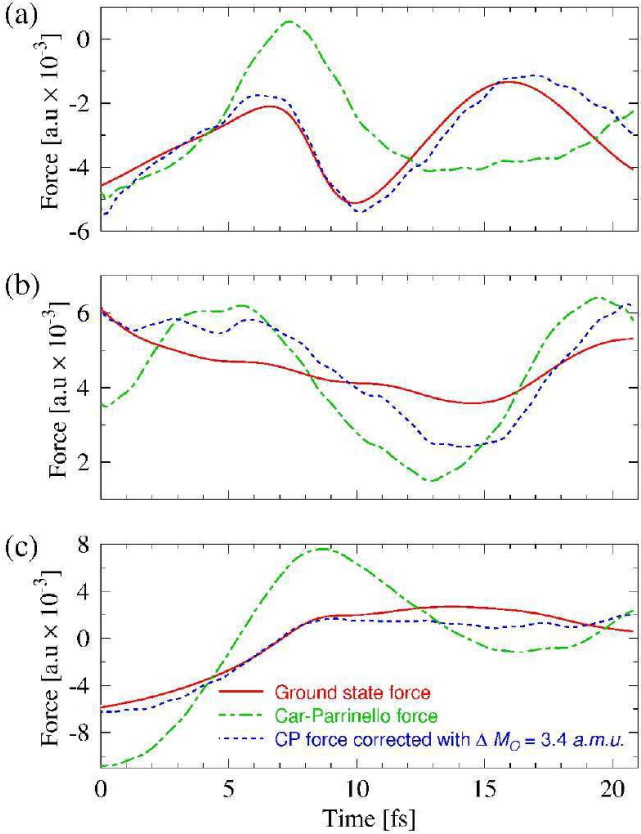,width=8.5cm} 
\caption{Some samples of force components on D$_2$O molecules along a $21$ fs segment of
CPMD trajectory. The Car-Parrinello force (green) and the Car-Parrinello force
once corrected with the best constant mass corrections (blue) is compared to the
ground state force at the same ionic positions. }
\label{fig:force2}
\end{figure}

In figure \ref{fig:force2} some sample force components on molecules are shown and compared
to the ground state force and the force once corrected using the average optimal mass corrections.

It is clear from Fig. \ref{fig:force1} and Fig. \ref{fig:force2} that the errors in CPMD forces 
due to the fictitious mass can not simply be described in terms of a constant correction to the masses of the ions.
In other words, the $\{\Delta F^{r \; \alpha}_I\}$  are quite large.

To get some perspective on the level of errors in the forces, 
these errors are compared in magnitude to those at varying
levels of convergence of the Kohn-Sham orbitals to their ground state
during a minimization using the method of direct inversion in
the iterative subspace (DIIS)\cite{diis}. 
Fifteen equally-spaced atomic `snapshot' configurations were extracted from the final $5.5$ 
picosecond production run with $\mu = 900$ a.u. and, starting with 
random wavefunctions, the Kohn-Sham energy was minimized. 
Convergence of this minimization was taken to mean that the
difference in energy between successive electronic iterations $\delta_{scf}$ 
had fallen lower than $10^{-13}$ a.u./atom. 
For each snapshot configuration all iterations for which $\delta_{scf}>10^{-3}$ a.u./atom
were discarded. At all other iterations the percentage r.m.s errors in the
forces $\Delta_{rms}$ relative to the fully converged forces were calculated.
\begin{figure}
\epsfig{figure=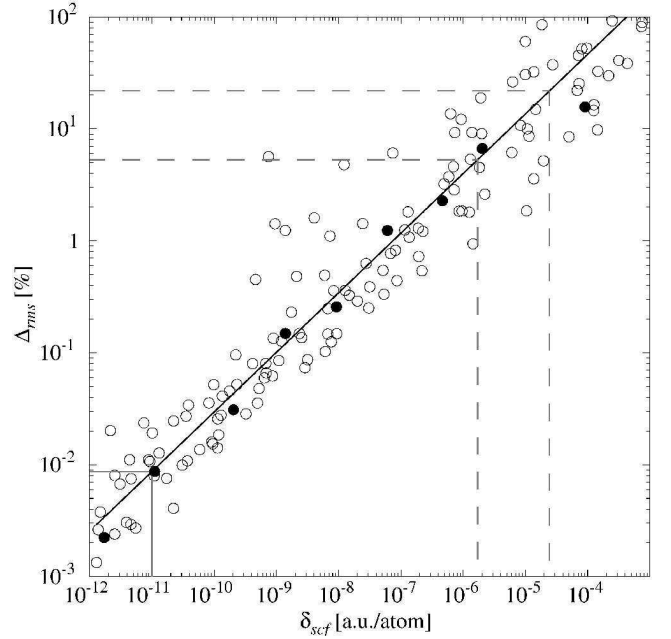,width=8.5cm} 
\caption{The r.m.s relative error $\Delta_{r.m.s}$ in the forces as a function of the
degree of convergence $\delta_{s.c.f}$ of the total energy during 
minimizations to the ground state with DIIS for $15$ 'snapshot' atomic configurations extracted
from the CPMD simulations. Each circle corresponds to a single self-consistency step in the minimization
of the wavefunctions for a single snapshot. The filled circles are the data from one full representative 
DIIS minimization and are included to illustrate the convergence characteristics for each snapshot.
The dashed lines indicate the value of $\Delta_{r.m.s.}$ from the mass corrected and uncorrected 
forces in the CPMD simulations and the level of convergence of the total energy that this would
correspond to. The grey line indicates the level of convergence that is typically used
in BOMD simulations of water ( $\delta_{s.c.f} = 10^{-11}$ a.u./atom ) and the value of $\Delta_{r.m.s.}$ that
this corresponds to.}
\label{fig:bomd}
\end{figure}
A line was then fit to the dependence of $\log_{10} \Delta_{rms}$ on
$\log_{10} \delta_{scf}$. This was done separately for the forces on the oxygen ions, the
deuterons and the water molecules. The comparison with the errors in CPMD is quantitatively
very similar in each
case and therefore only the results for the oxygen ions are plotted in
figure  \ref{fig:bomd}. Very similar results were also found when electronic minimization
was performed using a conjugate-gradients technique and therefore it is assumed that, for the 
purpose of the present discussion, the dependence of $\Delta_{r.m.s.}$ on $\delta_{s.c.f}$ could
 be assumed to be reasonably independent of the route taken to the electronic ground state.
For comparison, the values of $\Delta_{rms}$ for the forces from the
CPMD simulations both before and after the application of the rigid-ion mass corrections
are also shown in this figure~\ref{fig:bomd}.  The results suggest that, for the mass corrected (uncorrected)
forces, the degree of convergence required at each time step of a BOMD simulation in order to 
achieve an average error in the forces $\Delta_{r.m.s.}$ of the same magnitude is around
$\delta_{s.c.f} = 10^{-5}$ a.u./atom ($= 10^{-4}$ a.u./atom ). This is many orders of 
magnitude larger than the level of convergence that is generally used for water\cite{asthagiri_pre,
schwegler,kuo,vandevondele}. For example, in reference \onlinecite{kuo} BOMD simulations
were performed with a convergence of the energy at each step of $\delta_{s.c.f} = 10^{-11}$ a.u./atom.
It was reported that this simulation was about a factor of $4$ more expensive than a CPMD simulation
that used a time step of $4$ a.u. As illustrated in figure~\ref{fig:bomd} the errors in the forces are about
three orders of magnitude smaller than those in the CPMD simulation when $\delta_{s.c.f} = 10^{-11}$  a.u./atom\cite{force_comparison}.
A fair comparison between BOMD and CPMD should consider their relative efficiencies at the same
level of accuracy or their relative accuracies at the same level of efficiency. In general, this
comparison depends on the system under consideration and the level of accuracy (or efficiency) required.
No attempt will be made here to study this question in detail. However, it is illuminating briefly to
consider the relative efficiencies of BOMD and CPMD at the level of accuracy of a
CPMD simulation with $\mu=900$ a.u., and at the level of accuracy
of a BOMD simulations with $\delta_{s.c.f} = 10^{-11}$ a.u./atom .

It is assumed here that the efficiency of BOMD for ice is similar to that of water. 
Drawing on the results of Kuo {\em et al.}, it is therefore assumed that a BOMD simulation 
with $\delta_{s.c.f} = 10^{-11}$  is approximately three times slower than a CPMD simulation with 
$\mu = 900$ a.u. and $\Delta t = 6$ a.u.
Assuming that errors in the forces scale linearly with $\mu$, $\Delta_{r.m.s.}$ for the CPMD
simulation can be reduced to the level of the highly converged BOMD simulation by reducing $\mu$ by about
three orders of magnitude\cite{force_comparison}. The time step required to integrate the equations of motion for the orbitals
scales with the square-root of the fictitious mass, and therefore this reduction of $\mu$ requires a reduction 
of the time step (and therefore the speed of the simulation) by a factor of about $30$.
What this means is that, at this level of accuracy, BOMD is roughly an order of magnitude faster than CPMD.

If a much lower accuracy is sufficient, such as the level of accuracy in a CPMD simulation with $\mu = 900$ a.u., 
then a comparison with a BOMD simulation using a value of $\delta_{s.c.f}$ that is five or six orders 
of magnitude larger is sufficient. 
At this level of convergence, the BOMD simulation of Kuo {\em et al.} would be faster, but because $\delta_{scf}$ typically
reduces by an order of magnitude with each one or two electronic iterations, the speed up
may not be very great. It would depend on the average rate of convergence of the orbitals to the ground
state and on the quality of the extrapolation of the orbitals from previous time steps. It seems 
likely that the speeds of CPMD and BOMD simulations would be much closer in this case. However, a BOMD simulation 
at this level of accuracy would suffer from discontinuous forces - a problem that is not present in CPMD.

More detailed comparisons between CPMD and BOMD are clearly required. The point of the present discussion is 
simply that the accuracy (as judged by the quantity $\Delta_{rms}$) of a reasonably well-converged BOMD simulation 
far exceeds that of a CPMD simulation using standard values (a few hundred atomic units)\cite{large_mass} of the fictitious
mass. 
The argument that CPMD is more efficient than BOMD\cite{kuo} is frequently based on comparisons in which BOMD
is held to a higher standard of accuracy than CPMD.

\section{Phonon power spectrum}
\begin{figure} 
\epsfig{figure=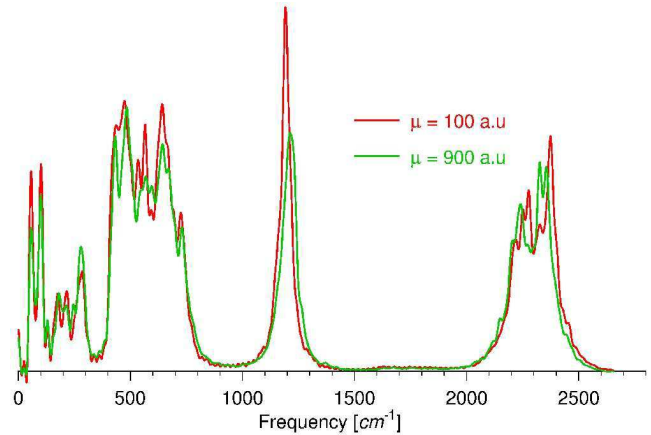,width=8.5cm} 
\caption{The mass-corrected phonon power spectra for simulations with $\mu=100$ a.u.
and $\mu=900$ a.u. }
\label{fig:phonon}
\end{figure}
As described in section ~\ref{section:details}, the phonon power spectra
 have been calculated for the simulations with $\mu = 100$ a.u.  and
$\mu = 900$ a.u. The results are plotted in figure \ref{fig:phonon}.
Different rescaled masses have been used in these two simulations and so the
coincidence of the main peaks in these power spectra clearly
vindicates the use of corrections to the masses of the ions. 
For MgO, it has already been demonstrated that the power spectrum is strongly
altered by the fictitious mass if mass corrections are not applied\cite{us}. In the present
work only the corrected power spectra are presented, but it should be clear that, if the
masses of the ions had not been rescaled {\em a priori}, the power spectrum of the simulation
with $\mu = 900$ a.u., in particular, would have been very different. 
Overall, the agreement between the two power spectra is reasonably good despite the fact that, 
as discussed in section ~\ref{section:forces}, the masses of the oxygen ions
vary considerably during the dynamics and that the mass correction that was
applied was simply a uniform constant correction to account for the average value
of the extra inertia of the ions due to the fictitious mass.
It may be that many time- or space-averaged properties are insensitive to $\mu$ as long as average
constant mass corrections are applied to the ions.
For example, the work of Grossman, Schwegler {\em et al.}\cite{grossman,schwegler} suggests 
that once adiabatic decoupling is achieved, the 
pair-distribution function of water is not very sensitive to the value of $\mu$\cite{footnote}.

\section{Discussion}
In the present work and in reference \onlinecite{us} it has been shown that the
inertia associated with the Kohn-Sham orbitals in CPMD can affect the motion
of the ions.  There are substantial differences between the CPMD forces
and the ground state forces at the same ionic positions. Reference \onlinecite{us}
focussed on two materials that exhibit extreme behaviour. Crystalline MgO
under pressure is an almost perfectly ionic material with a high quantum kinetic
energy associated with valence electrons that are strongly bound to the oxygen ions.
The $\mu$-dependent error for MgO is therefore very large but can almost perfectly
be corrected by applying a constant mass correction to the ions. Silicon, on the
other hand, is a material in which valence electrons are delocalized and have a small
quantum kinetic energy. $\mu$-dependent errors for silicon are therefore extremely small.

Here the focus has been on D$_2$O - an important system in which valence electrons possess a large
amount of quantum kinetic energy, but which is not perfectly ionic. It is demonstrated
that $\mu$-dependent errors are large, and while applying a constant mass correction
to the ions improves the forces considerably, systematic errors ($\Delta F^{r}$ ) remain that are
very large compared to the errors that would be present in a reasonably well-converged
BOMD simulation. These errors are highlighted in order to underline the importance
of caution when applying CPMD to situations in which substantial rearrangements
of the electronic orbitals are likely to occur. The non-rigid-ion part
of the  errors, $\Delta F^{r}$,  observed for crystalline D$_2$O are likely to be less serious than those
in liquid water, for example, where deviations of the electronic orbitals from their average
structure are likely to be larger. Furthermore, in studies of chemical
reactions and phase transitions deviations from rigid-ion behavior are likely to 
be more serious and it may not be safe to assume that dynamics and thermodynamics can
be corrected simply by using mass corrections for the ions. In such situations
it would appear safer to use a small value of $\mu$ or to use BOMD. 

In the past, the quality of a CPMD simulation has sometimes been judged according
to the degree of conservation of energy and the degree to which the adiabatic-decoupling
condition is maintained\cite{kuo}. 
In the present work it has been emphasized that adiabatic decouping
is not a decoupling of the ions from the orbitals but only 
a decoupling of the ultra high-frequency part of the orbital motion from the lower-frequency 
modes of the coupled orbital-ion system.
Perfect energy conservation and perfect adiabatic decoupling have both been maintained 
in the simulations of D$_2$O presented here, and yet the accuracy of the computed
forces is quite poor. 

Comparisons between BOMD and CPMD in this work have relied on calculating the
average magnitude of systematic errors, $\Delta F$,  (i.e. errors that do not average out on short time scales) 
in the forces on the ions. While there is no obvious alternative to using this quantity, and 
while the disparity in the magnitude of the errors between CPMD and BOMD appears large\cite{force_comparison}, 
it should be borne in mind that this is an extremely crude measure of the accuracy of a simulation.
The effects of the errors in the forces in CPMD and BOMD are likely to be different. Although an
analytic expression for the error in the forces in the case of CPMD is given by equation~\ref{eqn:error}
the effect on thermodynamic properties of the part of this error that does not reduce
to a mass correction, i.e. $\Delta \bf{F}^{r}$, is not obvious.

In BOMD, the errors in the forces lead to a systematic drift in the total energy due to a
systematic bias in the wavefunctions arising from their extrapolation from previous
time steps\cite{pulay}.
This energy drift can be a very useful judge of the accuracy of a simulation, however, 
a consequence of this is that there is also a drift in the temperature that, 
for long simulations, needs to be counter-balanced through the use of a thermostat for the ions. 
Apart from this obvious change in the temperature, the effects of errors in the BOMD forces on thermodynamic
properties is unknown, in general. 
Recent work by Pulay and Fogarasi has suggested counteracting the energy
drift by adding small corrections to the velocities of the ions at each time step\cite{pulay}.
This procedure might allow a BOMD simulation to be performed with a convergence criterion for the 
electronic structure that is less strict than commonly used criteria as it could be
both faster and with smaller average errors in the forces than are present in CPMD.
A disadvantage of performing poorly converged BOMD simulations is that the forces on the ions
would be discontinuous in time, and the simulations would not be time-reversible.
In conservative CPMD simulations forces are always continuous and simulations are always 
time-reversible regardless of the accuracy of the forces.

In CPMD the average kinetic energy of the ions can remain approximately constant throughout a simulation. 
In other words, while orbitals and ions are constantly exchanging energy and momentum in CPMD,
if the adiabatic decoupling condition is maintained there is no systematic net transfer 
of energy between them. Ions both gain momentum from and lose momentum to the orbitals in CPMD and the
physically meaningful total energy, which is the sum of the kinetic energy of the ions and the Kohn-Sham 
energy, fluctuates about a constant. 
The fluctuations in this energy are exactly mirrored by the fluctuations in the fictitious kinetic energy
because their sum is conserved. Therefore, the accuracy of CPMD can, in principle, be judged by the fluctuations 
in the FKE although how this should be done in practice is less obvious.  
As can be seen from figure \ref{fig:fke}, because care was taken to avoid
discontinuously accelerating the orbitals, only ionic-timescale fluctuations are visible in the
FKE. When the rigid-ion contribution to the FKE is subtracted out, the FKE that remains is much smaller
and with much smaller fluctuations, however, section \ref{section:forces} shows that the average remaining 
error in the forces is still orders of magnitude larger than would be found in a reasonably well converged
minimization of the orbitals to the electronic ground state. This is simply an illustration of the fact
that small fluctuations in the energy (which varies only to second order in $\delta \psi$) can result in much larger
fluctuations in the force (which varies to first order in $\delta \psi$ ).  It also suggests that it may be worth
examining the effect on {\em forces} rather than energies of the use of thermostats
to control the FKE\cite{blochl_and_parrinello,blochl2}.

Currently, {\em ab initio} molecular dynamics (AIMD) is hampered by two very important problems.
The first of these is the limitations on the time scales and length scales that
are accessible to simulations. This problem means that many properties and systems
are currently beyond the reach of AIMD. For simulations that can be performed, this problem
severely limits the precision with which thermodynamic properties are calculated.
The second major problem is that the density functionals that are currently in use are
not accurate enough for many applications, particularly in the study of chemical
reactions where a high level of accuracy is required on the energetics, and in
systems in which electron correlation plays a prominent role.
Both of these problems will gradually be reduced by technological, algorithmic and theoretical
innovations and therefore the relative importance of the kinds of errors described in the present 
work will increase.

It should be stressed that in the present work and in reference~\onlinecite{us}, 
apart from clear problems with vibrational properties due to increased ionic masses, 
and changes in the definition of temperature, thermal pressure, and 
other quantities that depend explicitly on the mass, no observable thermodynamic 
quantity has been demonstrated to deviate significantly from the value that would 
be obtained in a perfectly converged BOMD simulation. 
It may be that further testing will reveal that, in practice, many properties are rather 
insensitive to $\mu$ as long as constant mass corrections are applied to the ions. 
Clearly, given the magnitude of the errors demonstrated in the present work, 
and the lack of an obvious theoretical justification for CPMD when $\mu$ is large 
and the electronic structure deviates significantly from the rigid-ion limit, 
further testing is necessary.

\section{Conclusions}
What has been demonstrated in the present work is that
there are problems with some of the arguments that have been used in the past to theoretically justify CPMD.
The forces in CPMD differ from the ground state forces even when averaged on a femtosecond time scale
and even when constant mass corrections have been applied to the ions.
For commonly used values of the fictitious mass, $\mu$, the magnitude of these errors in the 
forces is large compared to the errors that would generally be tolerated in a BOMD simulation. 
How the errors in CPMD impact on thermodynamic properties is unknown.
The errors can systematically be reduced by reducing the value of $\mu$.

\begin{acknowledgments} 
The author is indebted to Sandro Scandolo for numerous useful discussions. The author
is also grateful to D. Prendergast, J. M. Herbert, M. Head-Gordon, D. Asthagiri, E. A. Carter, E. Schwegler, K. N. Kudin, and F. Giustino
 for useful discussions.
\end{acknowledgments}


\begin{references}
\bibitem{ortho} This orthogonality condition is more conventionally included
explicitly in the Lagrangian as a holonomic constraint, however, for the purposes
of clarity and convenience of notation it is made implicit in the functional derivatives 
throughout this paper.
\bibitem{almost}
There is an indirect dependence of the lowest frequency of the orbital subsystem
through the dependence of the ionic frequencies on $\mu$.
\bibitem{force_comparison}
The average error in the forces for $\delta_{scf}=10^{-11}$ obtained from the straight line fit to the
minimization results of Figure~\ref{fig:correction} is $\sim 0.0087 \%$. This is approximately $600$ ($2500$) times
smaller than the error in the mass corrected (uncorrected) forces in the CPMD simulation with $\mu = 900$ a.u.
\bibitem{footnote}
Corrections to temperature as described in Ref. ~\onlinecite{us} and in sections ~\ref{section:theory} and ~\ref{section:details}
of the present work were not applied in references ~\onlinecite{grossman} and ~\onlinecite{schwegler} and this complicates
the interpretation of their results, however, it seems unlikely that these temperature corrections are large enough
to qualitatively alter their findings.
\end{references}
\end{document}